\documentstyle[preprint,aps]{revtex}  
\tightenlines

\newcommand{\be}{\begin{equation}}
\newcommand{\ee}{\end{equation}}
\newcommand{\bea}{\begin{eqnarray}}
\newcommand{\eea}{\end{eqnarray}}
\newcommand{\ba}{\begin{array}}
\newcommand{\ea}{\end{array}}     
\newcommand{\nn}{\nonumber \\}
\newcommand{\half}{\frac{1}{2}}

\begin{document}
\draft 
\title{Particles and Strings in Degenerate Metric Spaces}
\author{Lu\'\i s A. Cabral and Victor O. Rivelles}
\address{Instituto de F\'{\i}sica, Universidade de S\~ao Paulo,\\
  Caixa Postal 66315, 05315-970 S\~ao Paulo, SP, Brazil\\
  e-mail: lacabral@fma.if.usp.br, rivelles@fma.if.usp.br}

\maketitle

\begin{abstract}
We consider relativistic and non-relativistic particles and 
strings in spaces (or space-times) with a degenerate metric. We show that
the resulting dynamics is described by a rich structure of
constraints. All constraints are classified and the dynamics depends
strongly on the parity of the difference between the dimension of the
space (or space-time) and the rank of the degenerate metric. For a
particular 
class of degenerate metrics we can identify the null eigenvectors of the
metric with its Killing vectors. We also give the first steps towards the
quantization of the non-relativistic particle using the Senjanovic
path integral quantization and the Batalin-Fradkin-Tyutin
conversion method. 
\end{abstract}

\vskip 2.5cm


\section{INTRODUCTION}
\label{sec1}

Singularities have a fundamental role in general relativity
\cite{sing}. However it is not clear how to incorporate them into a
quantum gravity theory. Surely singularities are an important feature
in quantum gravity as demonstrated by processes like Hawking
radiation. However our inability to follow the black hole evaporation
process till its very end shows our present limitations. In string
theory the situation is somehow ameliorated. The entropy of extreme
and near-extreme black holes, for instance,  can be 
computed by considering D-branes located near the horizon
\cite{entropy}. At weak coupling these D-branes are described by a
supersymmetric gauge theory for which it is possible to count the
relevant states for the entropy. Since the D-branes are located near
the horizon the singularities inside the
black hole have not yet been properly treated in string theory. There
are, however, some attempts in this direction \cite{ross}. 

The singularities which give rise to black holes manifest themselves
in the curvature tensor. The effect of this sort of singularity on the
propagation of quantum particles has recently been studied
\cite{horo-marolf}.  There is however another sort of singularity for
which the curvature tensor itself is not singular. They appear when
the metric tensor is degenerate and therefore has no inverse. If the
degeneracy occurs on a set of measure zero then the curvature remains
bounded and the topology of the space-time manifold can change
\cite{horowitz}. This sort of singularity is also compatible with the
causal structure of space-time \cite{chamblin}. Such singularities are
milder than curvature singularities and perhaps it should be easier to
handle them in a quantum gravity theory.

In general relativity degenerate metric appears in the Palatini
formulation. There we start with an action in terms of a tetrad
$e_\mu^a$ and a Lorentz connection $\omega_\mu^{ab}$. The action is $S
= \frac{1}{2} \int e^a e^b R^{c d} \epsilon_{abcd}$ where $e^a$ is the
tetrad one-form and $R = d\omega + \omega^2$ is the curvature
two-form. The space-time metric is then $g_{\mu\nu} = e_\mu^a e_\nu^b
\eta_{ab}$ where $\eta_{ab}$ is the tangent space Minkowski metric.
This action and the equations of motion which follow from it do not
depend on the inverse tetrad so they are well defined even when the
tetrad $e_{\mu}^a$ is degenerate. Solutions describing degenerate
tetrads have been found and a
relation between them and solutions of two-dimensional BF theories has
been established \cite{7}. In a less fundamental level
degenerate metrics also appear as induced metrics on null
hypersurfaces \cite{solowski}. In the four dimensional case the
induced degenerate metric is of rank two. 

Degenerate metrics or tetrads have also appeared in other
contexts. Ashtekar formulation of general 
relativity employs as phase space variables a complex potential
$A_{ai}$ and its electric field $E^{ai}$ in a complexified SO(3)
Yang-Mills theory \cite{ashtekar}. The usual metric interpretation of
general relativity requires that the spatial metric be proportional to
$E^{ai} E^{bi}$. Ashtekar formalism remains well defined even when
$E^{ai}$ is degenerate. Solutions presenting boundaries in which one
of the sides has a degenerate metric have been found \cite{9}. Also
many gravity theories in two and three 
dimensions can be formulated as a gauge theory of the Chern-Simons
type or BF type \cite{gauge}. All of them are valid when the
zweibein or dreibein are degenerate. It has also been suggested that there
would be a topological phase of quantum gravity in which the tetrad
vanishes and diffeomorphism invariance is unbroken \cite{witten-top}.

In the context of string theory it has been shown that strings can
propagate in degenerate metric backgrounds. When there
are only right or left movers the metric has a degeneracy of rank two 
\cite{strings}. It is also known that p-branes which behave as
instantons are described by a degenerate metric \cite{zaikov}. 
Also, tensionless strings \cite{lindstrom} and
tensionless D-branes \cite{unge} are governed by a dynamics that
involves a degenerate metric. 

All this indicates that degenerate metrics may have an important role
in quantum gravity. In fact, if we intend to incorporate degenerate
metrics in a quantum theory of gravity we should first understand its
classical effects. Some work has already been done in this
direction. At the classical level the main consequence of the 
degenerate metrics is to allow topology change of the space-time
manifold \cite{horowitz}. Since this is not observed in nature
there must exist a suppression mechanism for it, probably at the
quantum level. On the other side the behavior of matter in space-times
with a degenerate metric have not been properly explored. In the case
of topological field theories the metric appears nowhere in the action
so that the coupling to matter is trivial. Only global issues are
important in this case. However, when we consider matter fields with
canonical  
kinetic terms the coupling to gravity requires the introduction of
contravariant tensors fields. The simplest example is the action of
the scalar field which depends on the inverse metric. To treat these
cases an algorithm has been proposed which does not need any
contravariant tensor fields \cite{marolf}. However there is another
important class of matter which can be coupled to the metric
tensor but does not require contravariant tensor fields. It 
is constituted by particles and strings. The
action and field equations of particles and strings do not depend on
the inverse metric and are well defined even when the metric becomes
degenerate. 

In this paper we will consider the dynamics of particles
and strings in spaces or space-times with a degenerate
metric. We will analyze the structure of the constraints imposed by
the degenerate metric and find the effective degrees of freedom which
describe the motion of particles and strings. We will
first consider the non-relativistic particle in Section \ref{sec2}. A
complete analysis of the constraints will be done. It
will be developed in detail since the constraint analysis of the
subsequent sections will be very close to that of the non-relativistic
particle. In Section \ref{sec3} we will study the relativistic
particle and in Section \ref{sec4} strings will be
considered. In Section \ref{sec5} we will analyze the role of the
isometries of degenerate metrics and its relation to the analysis done
in the former sections. We will also consider a preliminary 
quantization of the non-relativistic particle in Section
\ref{sec6}. There we will use the path integral quantization due to
Senjanovic \cite{senjanovic}  and also the Abelian conversion method of
Batalin-Fradkin-Tyutin (BFT) \cite{BFT}. Finally Section \ref{sec7}
presents some remarks and conclusions.

\section{Non-Relativistic Particle}
\label{sec2}

Usually we interpret a degenerate metric space of dimension $D$ and
rank $r$ as a lower dimensional space with dimension $r$. However we
wish to consider a degenerate metric space in a broader sense. To
implement this consider a set of $D$ vectors in the lower dimensional
space $e_i^a$, $i=1,\dots,D$, $a=1,\dots,r$. 
From them we can build a degenerate metric $g_{ij} = e_i^a e_j^b
h_{ab}$, where $h_{ab}$ is an Euclidean metric, in such a way that
$g_{ij}$ has rank $r$. Indeed, if we consider a set of vectors
$U_\alpha$ in the D dimensional space then the equation $g_{ij}
U^j_\alpha = 0$, viewed as a 
linear system, has $D-r$ solutions for $U_\alpha$ since the rank of
$g_{ij}$ is $r$. These are the null eigenvectors of the metric. Notice
that only in the case that $e_i^a$ is locally integrable,
that is, if $e_i^a = \frac{\partial y^a}{\partial x^i}$, then it describes in
fact a $r$ dimensional space. 

Let us now consider a particle of mass $m$ moving in such a
space. Its Lagrangian is 
\be
\label{lagrangian}
L = \half m g_{ij} \dot{x}^i \dot{x}^j = \half m \dot{x}^a \dot{x}^b
h_{ab}, 
\ee 
where $\dot{x}^a = e^a_i \dot{x}^i$. The canonical momentum is 
\be
\label{momentum}
P_i = \frac{\partial L}{\partial \dot{x}^i} = m g_{ij} \dot{x}^j = m e_i^a
\dot{x}_a, 
\ee 
where $\dot{x}_a = h_{ab} \dot{x}^b$. Because the
metric is degenerate we can not invert this equation to find the
velocity $\dot{x}^i$ in terms of the momentum $P_i$. This means that the
system has constraints. If we consider the null eigenvectors $U_\alpha$ 
introduced above we find that 
\be 
P_i U^i_\alpha = m g_{ij} \dot{x}^j
U^i_\alpha = 0, 
\ee 
so that there are $D-r$ primary constraints 
\be
\label{constraints}
\phi_\alpha = P_i U^i_\alpha, \qquad \alpha = r+1 , \dots, D.  
\ee 
The range of the indice $\alpha$ was chosen in this way for future
convenience.

To find an explicit expression for $U_\alpha$ we consider,
without loss of generality, the
first $r$ components of Eq.(\ref{momentum}). They can be solved for the 
$r$ unknowns $\dot{x}^a$ as 
\be 
\dot{x}_a = \frac{1}{m} \tilde{e}^{i^\prime}_a P_{i^\prime},
\qquad i^\prime = 1, \dots, r, 
\ee 
where $\tilde{e}^{i^\prime}_a$ is
the inverse of $\tilde{e}^a_{i^\prime}$. Notice that
$\tilde{e}^a_{i^\prime}$ is a 
square matrix build up from the first $r\times r$ components of the
original $e_i^a$. The $D-r$ remaining equations in Eq.(\ref{momentum}) are
then 
\be 
P_\alpha = m e^a_\alpha \dot{x}_a  = e^a_\alpha
\tilde{e}_a^{i^\prime} P_{i^\prime}, 
\ee 
which give rise to the constraints found in Eq.(\ref{constraints}) 
\be
\label{phi_alpha}
\phi_\alpha = P_\alpha - e_\alpha^a \tilde{e}_a^{i^\prime}
P_{i^\prime}.  
\ee 
The explicit expression for $U_\alpha$ is then 
\be 
U_\alpha^i = \delta_\alpha^i -
e_\alpha^a \tilde{e}_a^{j^\prime} \delta_{j^\prime}^i.  
\ee
This structure is independent of the fact we have chosen the first
$r$ components of Eq.(\ref{momentum}) to solve for $\dot{x}_a$. Any other
choice would give equivalent results since we are solving algebraic
equations and the constraints are also algebraic.

The canonical Hamiltonian can then be found 
\be
\label{canonical_hamiltonian}
H_0 = P_i \dot{x}^i - L = \frac{1}{2m} \tilde{g}^{i^\prime j^\prime}
P_{i^\prime} P_{j^\prime}, 
\ee 
where $\tilde{g}^{i^\prime j^\prime} =
\tilde{e}_a^{i^\prime} \tilde{e}_b^{j^\prime} h^{ab}$ is a square
$r\times r$ matrix.  The Hamiltonian is then 
\be
\label{hamiltonian}
H = H_0 + \lambda^\alpha \phi_\alpha, 
\ee 
where $\lambda^\alpha$ are
the Lagrange multipliers which implement the constraints
$\phi_\alpha$.

The Poisson algebra of the constraints is found to be 
\bea
\label{constraint_algebra}
\{ \phi_\alpha, \phi_\beta \} &=& M_{\alpha \beta}, \nn M_{\alpha
  \beta} &=& M_{\alpha \beta}^{i^\prime} P_{i^\prime}, \nn M_{\alpha
  \beta}^{i^\prime} &=& \partial_{[\alpha} ( e_{\beta]}^a
\tilde{e}_a^{i\prime} ) - e_{[\alpha}^a \tilde{e}_a^{j^\prime}
\partial_{j^\prime} ( e_{\beta]}^b \tilde{e}_b^{i\prime} ),
\eea 
and the time evolution of the constraints give the consistency 
conditions 
\bea
\label{consistency}
& & N_\alpha + M_{\alpha \beta} \lambda^\beta = 0,  \\
\label{12a}
& & N_\alpha = N_\alpha^{i^\prime j^\prime} P_{i^\prime} P_{j^\prime},
\\ 
\label{12b} 
& & N_\alpha^{i^\prime j^\prime} = \frac{1}{2m} [ - \partial_\alpha
\tilde{g}^{i^\prime j^\prime} + e_\alpha^a \tilde{e}_a^{k^\prime}
\partial_{k^\prime} \tilde{g}^{i^\prime j^\prime} -
\tilde{g}^{k^\prime ( i^\prime} \partial_{k^\prime} ( e_\alpha^a
\tilde{e}_a^{j^\prime )} ) ].  
\eea 
It should be noticed that the above expression can not be written as a
linear combination of the constraints $\phi_\alpha$ since no
$P_\alpha$ appears in it. 

In order to perform the analysis of the consistency condition
Eq.(\ref{consistency}) it will be useful to consider first the situation
where we recover the usual interpretation for a degenerate metric as a
space of lower dimensionality. Then we will proceed to the general
case.

\subsection{Locally Integrable Case}

Let us first consider the situation where $e_i^a$ is locally
integrable, that 
is, it satisfies $\partial_{[i}e_{j]}^a = 0$ everywhere. Then we can
easily find that the line element reduces to $ds^2 = dx^a dx^b h_{ab}$
showing that the $x^a$ are the coordinates of an $r$-dimensional
space.

We then find from Eq.(\ref{constraint_algebra}) that $M_{\alpha \beta} =
0$ and from Eq.(\ref{12b}) that $N_\alpha = 0$ so that all
constraints are first class. The local symmetry 
generated by the constraints $\phi_\alpha$ is found in the usual way
\bea
\label{local_symmetries}
\delta x^{i^\prime} &=& \epsilon^\alpha \{ \phi_\alpha, x^{i^\prime}
\} = \epsilon^\alpha e_\alpha^a \tilde{e}_a^{i^\prime}, \nn 
\delta x^\alpha &= & - \epsilon^\alpha, 
\eea 
where $\epsilon^\alpha$ is the infinitesimal parameter. The Lagrangian
can then be written as 
\be 
\label{lagrangian1}
L = \half m \Delta^a \Delta^b h_{ab}, \qquad \Delta^a =
\tilde{e}^a_{i^\prime} \dot{x}^{i^\prime} + e_\alpha^a \dot{x}^\alpha.
\ee 
It is easily verified that this Lagrangian is invariant under
Eq.(\ref{local_symmetries}).

Now using the gauge freedom in Eq.(\ref{local_symmetries}) we can gauge
away all $x^\alpha$ so that we are left with $r$ degrees of freedom
$x^{i^\prime}$. The Lagrangian Eq.(\ref{lagrangian1}) then becomes 
\be 
L = \half m \tilde{g}_{i^\prime j^\prime} \dot{x}^{i^\prime}
\dot{x}^{j^\prime}, 
\ee 
describing a particle in geodesic motion in a
$r$-dimensional space with a non-degenerate metric $\tilde{g}_{i^\prime
  j^\prime}$.

\subsection{Non-integrable Case}

If $e_i^a$ is non-integrable then in general $M_{\alpha \beta}$ will
be non-vanishing. Then we have to solve Eq.(\ref{consistency})
for the Lagrange multipliers $\lambda^\alpha$. This involves inverting
Eq.(\ref{consistency}) and since $M_{\alpha \beta}$ is an
$(D-r)\times(D-r)$ antisymmetric matrix the existence of its inverse
will depend on $D-r$. So there are two possibilities depending whether
$N_\alpha$ vanishes or not.

\subsubsection{Non-vanishing $N_\alpha$}

Let us consider the case when $D-r$ is even. In this case $M_{\alpha
  \beta}$ in general has an inverse $M^{\alpha \beta}$ so that
  \footnote{We assume that $M$ has no null eigenvalues. If they
  exist further constraints will appear. The structure of the new
  constraints will depend explicitly on the form of $e_i^a$ and then 
  a general classification is no longer possible.}  
\be 
\lambda^\alpha = -  M^{\alpha \beta} N_\beta.  
\ee 
All constraints $\phi_\alpha$ are then second class. The
number of independent degrees of freedom in configuration space is then 
$\half(D+r)$.

The equations of motion for $x^i$ are 
\bea 
\dot{x}^{i^\prime} &=& \frac{1}{m} \tilde{g}^{i^\prime j^\prime}
P_{j^\prime} - e_\alpha^a 
\tilde{e}_a^{i^\prime} \lambda^\alpha, \nn \dot{x}^\alpha &=&
\lambda^\alpha, 
\eea 
while for $P^{i^\prime}$ we get 
\be
\dot{P}_{i^\prime} = - \frac{1}{2m} \partial_{i^\prime}
\tilde{g}^{j^\prime k^\prime} P_{j^\prime} P_{k^\prime} -
\partial_{i^\prime} (e_\alpha^a \tilde{e}_a^{j^\prime} ) P_{j^\prime}
\lambda^\alpha.  
\ee 
The Lagrangian can be written as 
\be 
L = P_{i^\prime} \dot{x}^{i^\prime} - \frac{1}{2m} \tilde{g}^{i^\prime
  j^\prime} P_{i^\prime} P_{j^\prime} + P_{i^\prime} e_\alpha^a
\tilde{e}_a^{i^\prime} \dot{x}^\alpha, 
\ee 
after using the fact that
$\phi_\alpha$ is a second class constraint. The first two terms of the
Lagrangian describes the motion of a particle in a $r$-dimensional
space with metric $\tilde{g}_{i^\prime j^\prime}$. However
$\tilde{g}$ depends on $x^\alpha$ but its momentum $P_{i^\prime}
e_\alpha^a \tilde{e}_a^{i^\prime}$ is not independent. Therefore the
phase space has $2r + (D-r) = D+r$ dimensions. There is a geodesic
motion in $r$ dimensions and in the remaining dimensions the
motion is constrained by the motion that takes place in
$r$ dimensions.

Let us now consider the situation when $D-r$ is odd. 
Now $M_{\alpha \beta}$ has no inverse since $\det M = 0$. Also the
number of constraints $\phi_\alpha$ is odd which means that either
there are first class constraints among the $\phi_\alpha$ or there are
new constraints. Indeed since Eq.(\ref{consistency}) can not
be solved for all $\lambda^\alpha$ it means that there is a relation
among $N_\alpha$ and $M_{\alpha \beta}$ not involving
$\lambda^\alpha$. Since Eq.(\ref{consistency}) comprises just
algebraic equations we can take anyone of them as the new constraint.
Let us choose the last component of Eq.(\ref{consistency}) as
the new constraint $\chi$. Then we can solve the first $D-r-1$
components of Eq.(\ref{consistency}) for the first $D-r-1$
components of $\lambda^\alpha$ 
\be
\label{lambda}
\lambda^{\beta^\prime} = - M^{\alpha^\prime \beta^\prime} (
 N_{\alpha^\prime} + M_{\alpha^\prime, D-r} \lambda^{D-r} ),
\qquad \alpha^\prime, \beta^\prime = 1, \dots, D-r-1, 
\ee 
assuming that $M_{\alpha^\prime \beta^\prime}$ has an inverse (see
footnote 1). The new constraint is then 
\be
\label{new_constraint}
\chi = \dot{\phi}_{D-r} =   N_{D-r} - M^{\alpha^\prime
  \beta^\prime} M_{\alpha^\prime, D-r} N_{\beta^\prime}, 
\ee 
where use was made of Eq.(\ref{lambda}). No dependence of
  $\lambda^{D-r}$ is left as it should be. 

We also find the Poisson algebra for the new constraint 
\be
\label{new_algebra}
\{ \phi_\alpha, \chi \} = \frac{\partial \chi}{\partial P_{i^\prime}}
\partial_{i^\prime} \phi_\alpha - \frac{\partial \phi_\alpha}{\partial
  P_\beta} \partial_\beta \chi - \frac{\partial \phi_\alpha}{\partial
  P_{i^\prime}} \partial_{i^\prime} \chi,
\ee 
which in general is non-vanishing indicating that the new constraint
is second class. The Hamiltonian is now 
\be 
H = H_0 + \lambda^\alpha \phi_\alpha + \mu \chi,
\ee 
where $\mu$ is a new Lagrange multiplier. The consistency
conditions are now 
\bea
\label{first}
\dot{\chi} &=& \{ \chi, H_0 \} + \lambda^{\alpha^\prime} \{ \chi,
\phi_{\alpha^\prime} \} + \lambda^{D-r} \{ \chi, \phi_{D-r} \} = 0, \\
\label{second}
\dot{\phi}_{\alpha^\prime} &=& N_{\alpha^\prime} +
\lambda^{\beta^\prime} M_{\alpha^\prime \beta^\prime} + \lambda^{D-r}
M_{\alpha^\prime D-r} + \mu \{ \phi_{\alpha^\prime},\chi \} = 0, \\
\label{third}
\dot{\phi}_{D-r} &=& N_{D-r} + \lambda^{\beta^\prime} M_{D-r 
  \beta^\prime} + \mu \{ \phi_{D-r}, \chi \} = 0.  
\eea 
As before, from Eq.(\ref{second}) we find $\lambda^{\alpha^\prime}$ 
\be
\lambda^{\beta^\prime} = - M^{\alpha^\prime \beta^\prime} (
N_{\alpha^\prime} + \lambda^{D-r} M_{\alpha^\prime D-r} + \mu \{
\phi_{\alpha^\prime}, \chi \} ).  
\ee 
Then Eq.(\ref{third}) can be rewritten as 
\be 
\chi + \mu ( M^{\alpha^\prime \beta^\prime} \{
\phi_{\alpha^\prime} , \chi \} M_{D-r \beta^\prime} + \{ \phi_{D-r},
\chi \} ) = 0, 
\ee 
which on the constraint surface gives $\mu = 0$. Finally,
  Eq.(\ref{first}) allows us to determine $\lambda^{D-r}$ 
\be
\lambda^{D-r} = - \frac{ \{ \chi, H_0 \} + M^{\alpha^\prime
  \beta^\prime} N_{\alpha^\prime} \{ \chi, \phi_{\beta^\prime } \} } { \{ 
  \chi, \phi_{D-r} \} - M^{\alpha^\prime \beta^\prime}
  M_{\alpha^\prime D-r} \{ \chi, \phi_{\beta^\prime} \} }.  
\ee 
Then all Lagrange multipliers are determined \footnote{This solution
  is good if $\{ \chi, \phi_{D-r} \} + M^{\alpha^\prime \beta^\prime}
  N_{\alpha^\prime} \not= 0$. It is very hard to handle this
  condition since it involves explicitly the inverse of
  $M_{\alpha^\prime \beta^\prime}$. In all cases we verified this
  condition is true but we do not have a general proof of that. If it
  fails then new constraints would be possible.} 
which means that all constraints are second class. Therefore we have 
$\frac{1}{2}(D+r-1)$ independent degrees of freedom in configuration
  space. The dynamics is 
similar to the former case. It describes the motion of a particle in
an $r$-dimensional metric space with metric $\tilde{g}$ with the
momentum being constrained by $\phi_\alpha$ and $\chi$ so that we have
a phase space with $2r + (D - r -1 )= D + r -1$ dimensions.

\subsubsection{Vanishing $N_\alpha$}

When solving Eq.(\ref{consistency}) another possibility has to be
taken into account. If the degenerate metric is such that
$N_\alpha^{i^\prime j^\prime}$ vanishes a little algebra shows that
\begin{equation}
\partial_{[\alpha} \tilde{e}_{k^\prime]}^c + e_\alpha^a
\tilde{e}_a^{l^\prime} \partial_{[ k^\prime} \tilde{e}_{l^\prime]}^c =
0,
\end{equation}
and
\begin{equation}
M_{\alpha\beta}^{i^\prime} = \left( \partial_{[\alpha} e_{\beta ]}^c -
  e_{[\alpha}^a e_{\beta ]}^b \tilde{e}_a^{j^\prime}
  \tilde{e}_b^{k^\prime} \partial_{j^\prime} \tilde{e}_{k^\prime}^c
  \right) \tilde{e}_c^{i^\prime}.
\end{equation}
Then we have to solve $M_{\alpha\beta} \lambda^\beta = 0$ with
$M_{\alpha\beta}$ given above.

Again we have to consider the cases $D-r$ even and odd separately. 
If $D-r$ is even in general $M_{\alpha\beta}$ has an inverse (see
footnote 1) and all
Lagrange multipliers can be determined $\lambda^\alpha = 0$. All
constraints are second class and in configuration space we find
$(D+r)/2$ degrees of freedom. The equations of motion for $x^i$ are
\begin{eqnarray}
\dot{x}^{i^\prime} &=& \frac{1}{m} \tilde{g}^{i^\prime j^\prime}
P_{j^\prime}, \nonumber \\
\dot{x}^\alpha &=& 0,
\end{eqnarray}
while for $P_{i^\prime}$ we obtain
\begin{equation}
\dot{P}_{i^\prime} = - \frac{1}{2m} \partial_{i^\prime}
\tilde{g}^{j^\prime k^\prime} P_{j^\prime} P_{k^\prime}.
\end{equation}
The independent degrees of freedom are then $x^{i^\prime}$ and
$P_{i^\prime}$. 

For $D-r$ odd  again $M_{\alpha\beta}$ has no inverse so some
constraints are first 
class or there must exist new constraints. Following the same
procedure as before we can solve
$M_{\alpha\beta} \lambda^\beta = 0$ for $D - r -1$ Lagrange
multipliers $\lambda^{\alpha^\prime} (\alpha^\prime, \beta^\prime = 1
\dots D-r-1)$ leaving $\lambda^{D-r}$ undetermined
\begin{equation}
\lambda^{\alpha^\prime} = - M^{\alpha^\prime \beta^\prime}
M_{\alpha^\prime D-r} \lambda^{D-r}.
\end{equation}
This means that there must exist one first class constraint which is
given by
\begin{equation}
\Gamma = \phi_{D-r} - M_{\alpha^\prime D-r} M^{\alpha^\prime
  \beta^\prime} \phi_{\beta^\prime}. 
\end{equation}
Therefore there are $D-r-1$ second class constraints
$\phi_{\alpha^\prime}$ and one first class constraint $\Gamma$ and
consistency gives no more constraints. Then we have $2D - (D - r -1)
-2 = D+r-1$ degrees of freedom in phase space.

\section{The Relativistic Particle}
\label{sec3}

We will work in a space-time with signature $(-,+,+, \dots)$.  To
avoid problems with the interpretation of a degenerate metric in a
time-like coordinate we will consider only degeneracy in space-like
surfaces. Then the world-line of the particle will be well defined. So we
will consider a degenerate metric of minimal rank of 2.

The action for a relativistic particle of mass $m$ in a gravitational
background is described in the first order formalism by

\begin{equation}
S = \int dt \left( \frac{1}{2e} g_{\mu\nu} \dot{x}^\mu\dot{x}^\nu
- \frac{e}{2} m^2 \right),
\end{equation}
where $e$ is the einbein. The canonical momenta to $e$ and $x^{\mu}$
are, respectively, 

\begin{equation}
\Pi_e = \frac{\partial L}{\partial \dot{e}} = 0, \qquad
P_\mu = \frac{\partial L}{\partial \dot{x}^\mu} =
e^{-1}g_{\mu\nu}\dot{x}^\nu.
\end{equation}
Assuming that the metric is not degenerate we find only one
constraint $\Pi_e=0$. The canonical Hamiltonian is
easily found to be 
\begin{equation}
H_c =\frac{e}{2} \left( P_\mu P_\nu g^{\mu\nu} + m^2 \right),
\end{equation}
and consistency leads to the usual constraint $P^2 + m^2 = 0$. Both
constraints are first class and the canonical Hamiltonian 
vanishes as a consequence of the reparametrization invariance of the
action.  

When the metric is degenerate we can not invert the velocities in
terms of the momenta. This means that new primary constraints will
appear. So let us consider again a metric of the form $ g_{\mu\nu} =
e^{a'}_\mu e^{b'}_\nu \eta_{a' b'}$ where 
$\eta_{a' b'} = diag(-1,+1, \dots , +1)$, and $\mu,\nu=0,1,\dots,D-1$,
$a', b' = 0, 1, \dots , r-1$. It is easily verified that the
determinant of this metric vanishes and that it has rank $r$. 
Let us split the range of the curved space indices
 into $\mu^\prime,\nu^\prime=0,1,\dots,r-1$ 
and $\underline{\mu},\underline{\nu}=r,\dots,D-1$. 
With this choice for the metric there is an additional primary 
constraint. It is given by 
\begin{equation}
\Phi_{\underline{\mu}} = P_{\underline{\mu}} - e^{a^\prime}_{\underline{\mu}}
\tilde{e}_{a^\prime}^{\nu^\prime} P_{\nu^\prime},
\end{equation}
and satisfies the Poisson bracket algebra
\begin{equation}
\left\{ \Phi_{\underline{\mu}} , \Phi_{\underline{\nu}} \right\} =
M_{\underline{\mu} \underline{\nu}},
\end{equation}
where
\begin{eqnarray}
M_{\underline{\mu} \underline{\nu}} &=&
P_{\lambda^\prime} \left(
\partial_{[ \underline{\mu} }
h^{\;\; \lambda^\prime}_{\underline{\nu} ]} -
h^{\;\; \rho^\prime}_{[ \underline{\mu} }
\partial_{\rho^\prime}
h^{\;\; \lambda^\prime}_{\underline{\nu} ]}
\right), \\
N_{\underline{\mu}} &=& e \tilde{P}_{a^\prime}\tilde{P}_{d^\prime}
\tilde{e}^{\chi^\prime}_{c^\prime}
\eta^{c^\prime d^\prime}
\left(
\partial_{[ \underline{\mu}}e^{a^\prime}_{\chi^\prime ]} -
h^{\lambda^\prime}_{\underline{\mu}}
\partial_{[ \lambda^\prime}e^{a^\prime}_{\chi^\prime ]}
\right), 
\end{eqnarray}
and $h_{\underline{\mu}}^{\lambda^\prime} =
e_{\underline{\mu}}^{a^\prime}
\tilde{e}_{a^\prime}^{\lambda^\prime}$. The canonical Hamiltonian is
given by  
\begin{equation}
H_{c} = \frac{e}{2} \left( \tilde{P}^2 + m^2 \right), 
\end{equation}
where
\begin{equation}
\tilde{P}^2 =
\tilde{g}^{{\mu^\prime}{\nu^\prime}} P_{\mu^\prime} P_{\nu^\prime}, \qquad
\tilde{g}^{{\mu^\prime}{\nu^\prime}} =
\eta^{{a^\prime}{b^\prime}}\tilde{e}^{\mu^\prime}_{a^\prime}
\tilde{e}^{\nu^\prime}_{b^\prime}.
\end{equation}
It follows that the the primary Hamiltonian is 
\begin{equation}
H_{p} = \frac{e}{2} \left( \tilde{P}^2 + m^2 \right) + \lambda \Pi_{e} +
\lambda^{\underline{\mu}}\Phi_{\underline{\mu}},
\end{equation}
where $\lambda$ and $\lambda^{\underline{\mu}}$ are Lagrange multipliers for
the primary constraints. 

The consistency of the primary constraints result in the secondary
constraint
\begin{equation}
\varphi =  - \frac{1}{2}\left( \tilde{P}^2 + m^2 \right), 
\label{Phi.mu}
\end{equation}
and the consistency condition
\begin{equation}
\label{47}
N_{\underline{\mu}} + \lambda^{\underline{\nu}} M_{\underline{\mu} \;
  \underline{\nu}} = 0. 
\end{equation}
Notice that the consistency of $\Pi_e$ results in a constraint  
analogous to usual case with the important difference that $\tilde{P}$
which appears in it and has a form which is similar to that of the
non-relativistic case.

Assuming that $M_{\underline{\mu} \underline{\nu}}$ and
  $N_{\underline\mu}$ do not vanish we can consider, as in the
  non-relativistic particle case, two situations  
  depending on the parity of $D-r$. 

In the case with $D-r$ even the inverse of $M_{\underline{\mu}
  \underline{\nu}}$ there exists (see footnote 1) and we solve 
Eq.(\ref{47}) for the Lagrange multiplier
\begin{equation}
\lambda^{\underline{\nu}} = - N_{\underline{\mu}}
M^{\underline{\mu}\underline{\nu}}. 
\end{equation}
There are then $D-r+1$ second class constraints, that is, an odd number
of them. This means that there must exist first class constraints among
them. To find them let us take the linear combination 
\begin{equation}
\phi = f \varphi + f^{\underline{\mu}} \phi_{\underline{\mu}},
\end{equation}
and impose that it is first class. We then find 
\begin{eqnarray}
\frac{2}{e} f^{\underline{\mu}}N_{\underline{\mu}} &=& 0, \\
-\frac{2}{e} f N_{\underline{\rho}} +
f^{\underline{\mu}}M_{\underline{\mu} \underline{\rho}} &=& 0.
\end{eqnarray}
The solution for these equations are
\begin{eqnarray}
f^{\underline{\mu}} &=& N_{\underline{\rho}}
M^{\underline{\mu} \underline{\rho}}, \\
f & = & \frac{e}{2}.
\end{eqnarray}
Then the first class constraint is given by
\begin{equation}
\phi = \frac{e}{2}\varphi +
M^{\underline{\mu} \underline{\rho}}
N_{\underline{\rho}}
\Phi_{\underline{\mu}}.
\end{equation}
Therefore there are $D-r$ second class constraints
$\Phi_{\underline{\mu}^\prime}$ 
and one first class constraint $\phi$. This gives $D+r-2$ degrees of
freedom in phase space. 

In the case of $D-r$ odd there is no inverse for
$M_{\underline{\mu\nu}}$. Proceeding as in 
the non-relativistic case we select the last component
Eq.(\ref{47}) 
using the notation $\underline{\mu} = (\underline{\mu}^\prime ,
\underline{D-r}),\underline{\mu}^\prime = 1,...,D-r-1 $. We can then
solve for $\lambda^{\underline{\nu}^\prime}$ finding
\be
\lambda^{\underline{\nu}^\prime} =
-M^{\underline{\mu}^\prime \underline{\nu}'} \left(
N_{\underline{\mu}^\prime}  +
M_{\underline{\mu}' \underline{D-r}} \lambda^{\underline{D-r}}
\right),
\ee
assuming that $M_{\underline{\mu^\prime \nu^\prime}}$ has an inverse
(see footnote 1). The component $\underline{D-r}$ of
Eq(\ref{47}) gives a new constraint  
\be
\chi = N_{\underline{D-r}}  -
M_{\underline{D-r} \underline{\beta}'}
M^{\underline{\alpha}^\prime \underline{\beta}'}
N_{\underline{\alpha}^\prime}.
\ee

The set of constraints is now $\Pi_e, \varphi$ and $\chi$. From the
time evolution of $\chi$ we get 
\be
\label{before}
\lambda^{\underline{D-r}} = - \frac{ \frac{e}{2} \left\{ \chi ,
    \varphi \right\} 
+ M^{\underline{\rho}^\prime \underline{\mu}'} N_{\underline{\rho}'}
\left\{ \chi, \Phi_{\underline{\mu}'} \right\} }{ \left( \left\{ \chi,
\Phi_{\underline{D-r}} \right\} - M^{\underline{\rho}^\prime
\underline{\mu}'} M_{\underline{\rho}^\prime \underline{D-r}} \left\{ \chi ,
\Phi_{\underline{\mu}'} \right\} \right)}.
\ee
To be sure that the denominator of Eq.(\ref{before}) does no vanish
consider the null eigenvector of $M_{\underline{\mu} \underline{\nu}}$.
It can easily be found to be
\be
\label{Vmu}
V^{\underline{\mu}} = \delta^{\underline{\mu}}_{\underline{D-r}} -
M_{\underline{D-r} \; \underline{\beta}'}
M^{\underline{\beta}' \; \underline{\nu}'}
\delta^{\underline{\mu}}_{\underline{\nu}'}.
\ee
We then find that the denominator of Eq.(\ref{before}) can be written as 
$V^{\underline{\mu}} \{ \chi , \Phi_{\underline{\mu}}\}$ and it is
non-vanishing. 

We have then found that the Lagrange multipliers
$\lambda^{\underline{\nu}}$ and $\lambda^{\underline{D-r}}$ are
determined. Therefore the constraints $\Phi_{\underline{\mu}}$ are
second class. We 
have now to find out which class the constraint $\chi$ belongs
to. Following the same steps as in the non-relativistic case we
consider the extended Hamiltonian 
\be
H = \frac{e}{2} \varphi + \lambda^{\underline{\mu}}\Phi_{\underline{\mu}} +
\mu \chi.
\ee
Using it to evaluate the time evolution of the constraints we find that 
\bea
\label{phiponto}
\dot{\varphi}& =& \lambda^{\underline{\mu}} \{ \varphi ,
\Phi_{\underline{\mu}} \} + \mu \{ \varphi , \chi \} \nonumber\\
&=& \frac{-2}{e} \lambda^{\underline{\mu}'} N_{\underline{\mu}'}
-\frac{2}{e} \lambda^{\underline{D-r}} N_{\underline{D-r}} +
\mu \{ \varphi , \chi \} = 0,
\eea
\be
\label{chi.dot}
\dot{\chi} = \frac{e}{2} \{ \chi , \varphi \} + \lambda^{\underline{\mu}}
\{ \chi , \Phi_{\underline{\mu}} \} = 0,
\ee
\be
\label{phi.mu'}
\dot{\Phi}_{\underline{\mu}'} =
N_{\underline{\mu}'} + \lambda^{\underline{\nu}'} M_{\underline{\mu}' \;
\underline{\nu}' } + \lambda^{\underline{D-r}} M_{\underline{\mu}' \;
\underline{D-r} } + \mu \{ \Phi_{\underline{\mu}'} , \chi \} = 0,
\ee
\be
\label{phi.dot}
\dot{\Phi}_{\underline{D-r}} = N_{\underline{D-r}} +
\lambda^{\underline{\nu}'} M_{\underline{D-r}' \; \underline{\nu}' } +
\mu \{ \Phi_{\underline{D-r}} , \chi \} = 0.
\ee
From Eq.(\ref{phi.mu'}) we get 
\be
\label{lambda_a'}
\lambda^{\underline{\alpha}'} = -M^{\underline{\alpha}' \underline{\mu}' }
\left( N_{\underline{\mu}'} + \lambda^{\underline{D-r}}M_{\underline{\mu}'
\underline{D-r}} + \mu\{ \Phi_{\underline{\mu}'} , \chi \} \right).
\ee
From Eq.(\ref{phi.dot}) we get $\mu=0$ and from Eq.(\ref{chi.dot}) we
find again the result already found in Eq.(\ref{before}). Finally
Eq.(\ref{phi.dot}) vanishes identically. 
Therefore all Lagrange multipliers associated to the constraints  $\chi$
and $\Phi_{\underline{\mu}}$ are determined showing that they are
second class. We also have that 
\be
\{ \varphi , \Phi_{\underline{\mu}} \} = - \frac{2}{e}N_{\underline{\mu}}
\neq 0, 
\ee
so that $\varphi$ is also second class. This gives an odd number of
second class constraints for $D-r$ odd. This implies then that there
must exist a combination of them which is first class. 

To find out the first class combination let us denote the set of
second class constraints as  
\be
\Gamma_{(\xi)} = \left( \varphi , \chi , \Phi_{\underline{\mu}} \right) =
 \left( \Gamma_{(A)} , \Gamma_{(B)} , \Gamma_{(\underline{\mu})} \right),
\ee
and the matrix of the Poisson brackets of $\Gamma_{(\xi)}$ as
$\gamma_{(\xi) \; (\eta)} \; =  \{ \Gamma_{(\xi)} , \Gamma_{(\eta)} \}$
so that $\det \gamma = 0$. The null eigenvector of $\gamma$, which
satisfies $W^{(\xi)}\gamma_{(\xi) \; (\eta)} = 0$, is then 
\be
W^{(\xi)} = \delta^{(\xi)}_{(D-r)} - \gamma_{(D-r) \; (\eta')}
\gamma^{(\eta') \; (\zeta')} \delta^{(\xi)}_{(\zeta')}. 
\ee
By imposing that the following linear combination of $\Gamma_{(\xi)}$
\be
\xi = W^{(\xi)} \Gamma_{(\xi)},
\ee 
is first class results in 
\be
\label{xi}
\xi = \Phi_{\underline{D-r}} - \left\{ \Phi_{\underline{D-r}} ,
\Gamma_{(\eta')} \right\} \gamma^{(\eta') \; (\xi')} \Gamma_{(\xi')}.
\ee
We then have the second class constraints $\varphi, \chi$ and
$\Phi_{\underline{\mu}}$ and one first class constraint $\xi$ so that
there are $D+r-5$ degrees of freedom in phase space. 

\section{Strings}
\label{sec4}

The action for a string in a gravitational background $G_{\mu\nu}(x)$
in $D$ dimensions is given by 
\begin{equation}
\label{eqstraction1}
S = - \frac{1}{4\pi\alpha'}\int d\sigma d\tau \sqrt{-h}
h^{(a)(b)}(\sigma, \tau)G_{\mu\nu}(x(\sigma,\tau))
\partial_{(a)}x^{\mu}\partial_{(b)}x^{\nu},
\end{equation}
where $h_{(a)(b)}$ is the world-sheet metric, $\mu,\nu = 0,...,D-1$
and $(a), (b)= 0,1$.  The equations of motion are 
\begin{equation}
\label{eqxmu1}
\frac{1}{\sqrt{-h}}\partial_{(a)}(\sqrt{-h}h^{(a)(b)}\partial_{(b)}x^{\mu})
G_{\lambda\mu} +
\Gamma_{\mu\nu\lambda}h^{(a)(b)}\partial_{(a)}x^{\mu}\partial_{(b)}x^{\nu}
=0, 
\end{equation}
\begin{equation}
\label{eqhab1}
T_{(a)(b)} = G_{\mu\nu}(\partial_{(a)}x^{\mu}\partial_{(b)}x^{\nu} -
\frac{1}{2}h_{(a)(b)}h^{(a ')(b ')}\partial_{(a ')}x^{\mu}
\partial_{(b')}x^{\nu}) = 0,
\end{equation}
and do not depend on the inverse of $G_{\mu\nu}$. Therefore the action
and equations of motion are well defined even in the case of
degenerate metrics. 

In the orthogonal gauge, $h_{ab}=\eta_{ab}$, Eqs.(\ref{eqxmu1},
\ref{eqhab1}) reduce to 
\begin{equation}
\label{eqxmu2}  
G_{\lambda\mu}{\hat{\partial}}_{+}{\hat{\partial}}_{-}x^{\mu} +
\Gamma_{\mu\nu\lambda}{\hat{\partial}}_{+}x^{\mu}{\hat{\partial}}_{-}x^{\nu}
 = 0,
\end{equation}
\begin{equation}
\label{eqhab2} 
T_{\pm\pm} = G_{\mu\nu}{\hat{\partial}}_{\pm}x^{\mu}
{\hat{\partial}}_{\pm}x^{\nu} = 0,
\end{equation}
where ${\hat{\partial}}_{\pm} = \partial_{\tau} \pm \partial_{\sigma}$,
$\partial_{\tau} = \partial_{0}$ and $\partial_{\sigma} =
\partial_{1}$. 

As we can notice there are chiral solutions for the equations
of motion Eqs.(\ref{eqxmu2}, \ref{eqhab2}) in the form
${\hat{\partial}}_{+}x^{\mu} = 0$ or ${\hat{\partial}}_{-}x^{\mu} = 0$
which are independent of the metric $G_{\mu\nu}$
\cite{strings}. However we know that the choice of the conformal
gauge still leaves a residual symmetry which can be fixed by the gauge
choice $ x^{+} \equiv \frac{1}{\sqrt{2}}(x^{0} + x^{D - 1}) =
(x_{0}^{+} + P\tau)$, where $P$ is a constant. In this gauge the
string coordinate 
$x^+$ is no longer chiral, that is  ${\hat{\partial}}_{+}x^{+} \neq 0
$. If we insist in having a chiral solution for the other string
coordinates we find that only certain gravitational backgrounds admit
such solutions \cite{strings}. They are given by 
\bea
\label{solnaodeg}
G_{--} &=& \tilde g_{--}, \qquad
G_{-i} = \tilde g_{-i}, \qquad G_{+j} = \tilde g_{ij}, \\
G_{+i} &=& \partial_{i}\tilde{h}, \qquad
G_{+-} = \partial_{-}\tilde{ h}, \qquad G_{++} = 0,
\eea
where $\tilde{g}_{--}, \tilde{g}_{-i}, \tilde{g}_{ij}$ and
$\tilde{h}$ are functions of $x^-$ and $x^i$. Among these allowed
gravitational backgrounds we can find some which give rise to a
degenerate metric. The most general form for the metric is given by
\cite{strings}
\begin{eqnarray}
G_{+-}&=& \partial_{-}\tilde{h},  \qquad G_{++} \; =
\; 0, \; \qquad G_{-i} \; = \; [a_{i}x^{+} + g_{i}(x^{i})]G_{+-}, 
\nonumber\\ & & \\
G_{+i}&=& a_{i}\tilde{h} + \partial_{i} \tilde{h}, \qquad  G_{--} \;
= \; 0, \qquad G_{ij} \; = \; \frac{G_{+(i}G_{-j)}}{G_{+-}}, 
\nonumber 
\end{eqnarray}
where $a_i$ are constants and $g_i$ are functions of $x^j$. We
then find that strings satisfying a chiral condition can propagate in a
degenerate gravitational background. 

We now develop the Hamiltonian formalism in order to find the
constraints when the metric in Eq.(\ref{eqstraction1}) is
degenerated. Introducing the string tension $T = 1/(2\pi{\alpha}^{'})$
we can find the momenta densities 
\begin{equation}
{\cal P}_{(a)(b)} = \frac{\partial{\it L}}{\partial{\dot{h}}^{(a)(b)}}
= 0, 
\end{equation}
\begin{equation}
\label{str:Pmu}
 {\cal P}_{\mu} = \frac{\partial{\it L}}{\partial{\dot{x}}^{\mu}} =
-T \sqrt{-h}h^{0(b)}\partial_{(b)}x^{\nu}G_{\mu\nu}(x).
\end{equation}
When the metric is non-degenerate the primary constraints are 
\begin{equation}
\label{Phiabndeg}
\Phi_{(a)(b)} = {\cal P}_{(a)(b)},
\end{equation}
and the primary Hamiltonian is
\begin{equation}
H = \int_{0}^{2\pi}d\sigma ({\cal H}_{c} +
\lambda^{(a)(b)}\Phi_{(a)(b)}),
\end{equation}
where $\lambda^{(a)(b)}$ are Lagrange multipliers and ${\cal H}_{c}$ is
the canonical Hamiltonian
\begin{equation}
\label{85}
{\cal H}_c = - \frac{1}{2h^{0 0}} \left[ \frac{(-h)^{-1/2}}{T} P_\mu
 P_\nu G^{\mu\nu} + 2 h^{01} P_\mu x^{'\mu} + T (-h)^{-1/2} x^{'\mu}
 x^{'\nu} G_{\mu\nu} \right].
\end{equation}
Consistency of the primary constraints leads to the secondary
constraints 
\be
\label{str:phi1}
\varphi_{1} = \frac{1}{T}{\it P}_{\mu}x^{'\mu},
\ee
\be
\label{str:phi2}
\varphi_{2} = \frac{1}{2}\left( \frac{1}{T^{2}}{\it P}_{\mu} 
{\it P}_{\nu} G^{\mu\nu} + x^{'\mu}x^{'\nu}G_{\mu\nu} \right). 
\ee
The secondary constraints do not generate any new constraints. As usual 
all constraints are first class and close the Virasoro algebra.

Let us now consider a degenerate metric of rank $r$ which can be set
in the form $G_{\mu\nu} = e^{a'}_{\mu}e^{b'}_{\nu}\eta_{a'b'}$ with
$a',b' =0, 1 \dots r-1$. As in the relativistic particle case we will
consider only degeneracy in the space-like directions. The primary
constraint Eq.(\ref{Phiabndeg}) is still 
the same but the momenta Eq.(\ref{str:Pmu}) can no longer be inverted
for the velocities. Following the development of the relativistic
particle we split the space-time indices $\mu,\nu$ according to
$\mu',\nu' = 0,1,...,r-1$ and
$\underline{\mu},\underline{\nu}=r,...,D-1$. This allows to define the
inverse of $e_{\mu^\prime}^{a^\prime}$ which is a square matrix as
$\tilde{e}_{a^\prime}^{\mu^\prime}$.  

Then, besides the constraint Eq.(\ref{Phiabndeg}) there are further
primary constraints which can be written as 
\be
\label{str:Phiumu}
\Phi_{\underline{\mu}} = {\cal P}_{\underline{\mu}} -
e^{a'}_{\underline{\mu}}\tilde{e}^{\nu'}_{a'}{\cal P}_{\nu'},
\ee
where
\bea
\label{Pmu}
{\cal P}_{\mu} &=& \Delta_{a'}e^{a'}_{\mu} , \nonumber\\
\Delta_{a'} &=& M^{0(b)}\partial_{(b)}x^{\nu}e^{b'}_{\nu}\eta_{a'b'}.
\eea
Notice that we can write this constraint as 
\bea
\Phi_{\underline{\mu}} &=& {\cal U}^{\lambda}_{\underline{\mu}}{\cal
  P}_{\lambda},  \\ 
{\cal U}^{\lambda}_{\underline{\mu}} &=&
\delta^{\lambda}_{\underline{\mu}} - 
e^{a'}_{\underline{\mu}}\tilde{e}^{\nu'}_{a'}\delta^{\lambda}_{\nu'},
\eea
showing that the metric $G_{\mu\nu}$ has $D-r$ null eigenvectors such
that ${\cal  U}^{\lambda}_{\underline{\mu}} G_{\lambda\nu} = 0$. 

The canonical Hamiltonian Eq.(\ref{85}) can then be rewritten as 
\be
\label{HamCanonArb}
{\mathcal H}_{c} = -\frac{1}{2h^{00}}\left[ \frac{(-h)^{-1/2}}{T}
\tilde{\mathcal P}^2 + 2h^{01}{\it P}_{\mu}x^{'\mu} +
T{(-h)}^{-1/2}x^{'2} \right],
\ee
where
\bea
\label{P2x'2}
\tilde{\mathcal P}^2 &=& \tilde{G}^{\mu' \nu'} {\mathcal P}_{\mu'} 
{\mathcal P}_{\nu'} = \eta^{a' c'} \tilde{e}^{\mu'}_{a'}
\tilde{e}^{\nu'}_{c'} {\mathcal P}_{\mu'} {\mathcal P}_{\nu'},
\nonumber \\
x'^{2} &=& x'^{\mu} x'^{\nu} G_{\mu \nu} = x'^{\mu} x'^{\nu}
e^{a'}_{\mu} e^{b'}_{\nu} \eta_{a' b'}.
\eea
We now build the primary Hamiltonian which will be used to find the
consistency conditions for the primary constraints. It is given by 
\be
\label{HamPrimArb}
H_p = \int \left( {\mathcal H}_c + \lambda^{(a)(b)}{\mathcal P}_{(a)(b)} +
\lambda^{\underline{{\nu}}} \Phi_{\underline{{\nu}}} \right) d\sigma',
\ee
where $\lambda^{(a)(b)}$ and $\lambda^{\underline{{\nu}}}$ are the Lagrange
multipliers. We then obtain the secondary constraints
\be
\label{CtrPhi1}
\varphi_{1} = \frac{1}{T} {\mathcal P}_{\mu} x^{'\mu},
\ee
\be
\label{CtrPhi2}
\varphi_{2} = \frac{1}{2} \left( \frac{\tilde{\mathcal P}^2}{T^2}
+ x^{'2} \right).
\ee 
This allows the canonical Hamiltonian Eq.(\ref{HamCanonArb}) to be
rewritten as 
\be
\label{HamCan-Ctr}
{\mathcal H}_c = - \frac{T}{ h^{00}} \left[ (-h)^{-1/2} \varphi_2 +
 h^{01} \varphi_1 \right].
\ee

Also the constraint algebra can be obtained. For $ \varphi_1$ and
$\varphi_2$ we obtain the usual Virasoro algebra
\be
\label{PBphi1phi1}
\left\{ \varphi_1(\sigma) , \varphi_1(\sigma') \right\} =
\left[ \varphi_1(\sigma) + \varphi_1(\sigma') \right] \partial_{\sigma}
\delta(\sigma - \sigma'),
\ee
\be
\label{PBphi1phi2}
\left\{ \varphi_1(\sigma) , \varphi_2(\sigma') \right\} =
\left[ \varphi_2(\sigma) + \varphi_2(\sigma') \right] \partial_{\sigma}
\delta(\sigma - \sigma'),
\ee
\be
\label{PBphi2phi2}
\left\{ \varphi_2(\sigma) , \varphi_2(\sigma') \right\} =
\left[ \varphi_1(\sigma) + \varphi_1(\sigma') \right] \partial_{\sigma}
\delta(\sigma - \sigma'),
\ee
while for $\Phi_{\underline{\mu}}$ we get
\be
\label{PBPhimuPhinu}
\{ \Phi_{\underline{\mu}}(\sigma) , \Phi_{\underline {\nu}}(\sigma') \} =
{\mathcal M}_{\underline{\mu} \; \underline{\nu}}(\sigma)
\delta(\sigma - \sigma'),
\ee
\be
\label{PBPhimuPhi1}
\{ \Phi_{\underline{\mu}}(\sigma) , \varphi_1(\sigma') \} =
- \Phi_{\underline{\mu}}(\sigma') \partial_{\sigma'}\delta(\sigma - \sigma'),
\ee
\be
\label{PBPhimuPhi2:2}
\{ \Phi_{\underline{\mu}}(\sigma) , \varphi_2(\sigma') \} =
n_{\underline{\mu}}(\sigma) \delta(\sigma - \sigma'), 
\ee
where
\bea
\label{StrMmunu}
{\mathcal M}_{\underline{\mu} \; \underline{\nu}} &=& {\mathcal P}_{\lambda'}
\tilde{e}^{\lambda'}_{b'}
\partial_{[\underline{\mu}}e^{b'}_{\underline{\nu}]}
(\sigma) + {\mathcal P}_{\rho'} e^{b'}_{[\underline{\nu}}
\partial_{\underline{\mu}]}\tilde{e}^{\rho'}_{b'}(\sigma) \nonumber\\
& & + {\mathcal P}_{\lambda'}\partial_{\rho'}\tilde{e}^{\lambda'}_{a'}
e^{a'}_{[\underline{\mu}}e^{b'}_{\underline{\nu}]}
\tilde{e}^{\rho'}_{b'}(\sigma) 
+ {\mathcal P}_{\lambda'} \tilde{e}^{\lambda'}_{a'}\tilde{e}^{\rho'}_{b'}
e^{b'}_{[\underline{\nu}}\partial_{\rho'}e^{a'}_{\underline{\mu}]}(\sigma),
\eea
\bea
\label{104}
n_{\underline{\mu}}(\sigma) &=& 2 \tilde{\mathcal P}_{a'}
\tilde{\mathcal P}_{d'} \eta^{c'd'}\tilde{e}^{\lambda'}_{c'}(\sigma)
\left[
\partial_{\left[\underline{\mu}\right.}e^{a'}_{\left.\lambda'\right]}(\sigma)
- e^{b'}_{\underline{\mu}}\tilde{e}^{\rho'}_{b'}
\partial_{[\rho'}e^{a'}_{\lambda']}(\sigma) \right] \nonumber\\
& & - U^{\beta}_{\underline{\mu}}\partial_{\beta}G_{\lambda \rho}
x^{'\lambda}x^{'\rho}(\sigma) - 2 \partial_{\sigma}U^{\nu}_{\underline{\mu}}
G_{\lambda \nu} x^{'\lambda}(\sigma). 
\eea

Consistency of $\Phi_{\underline{\mu}}$ then requires
\be
\label{Phi.mu-2}
{\cal N}_{\underline{{\nu}}}(\sigma) + \lambda^{\underline{{\mu}}}
{\cal M}_{\underline{{\mu}} \; \underline{{\nu}}}(\sigma) = 0,
\ee
where ${\cal N}_{\underline{\mu}}(\sigma) = c(\sigma)
n_{\underline{\mu}}(\sigma)$ and $c(\sigma) = -T/(2 \sqrt{-h} h^{00})$. 
This result is similar to those of the non-relativistic and relativistic
particles. Again we have to consider separately the cases in which
$D-r$ is even and odd. 

Let us assume that ${\cal N}_{\underline{\mu}} \not= 0$. 
For $D-r$ even there exists the inverse of ${\cal M}_{\underline{\mu}
  \; \underline{\nu}}$ (see footnote 1) 
and Eq.(\ref{Phi.mu-2}) can be solved for the Lagrange multiplier 
\be
\lambda^{\underline{\nu}} = - {\cal
M}^{\underline{\mu} \;
\underline{\nu}} {\cal N}_{\underline{\mu}}.
\ee
We then have a set with an odd number of second class constraints
$\Phi_{\underline{\mu}}$ 
and $\varphi_2$. We then take a linear combination of these 
constraints to select the first class constraint which is mixed with the
second class ones. We find that it is given by
\be
\label{phi1stclass}
\phi = c \varphi_2 +
{\cal N}_{\underline{\rho}} {\cal M}^{\underline{\rho} \; \underline{\mu}}
\Phi_{\underline{\mu}}. 
\ee
Then we have an even number of second class constraints
$\Phi_{\underline{\mu}}$ and the first class constraints $\phi,
\varphi_1$ and ${\cal P}_{(a)(b)}$ resulting in $D+r-4$ degrees of
freedom in phase space. 


For $D-r$ odd we follow the same steps as in the former cases. Since
the determinant of ${\cal M}_{\underline{\mu} \underline{\nu}}$
vanishes there is a null eigenvector ${\cal V}^{\underline{\mu}}$
such that 
${\cal V}^{\underline{\mu}} {\cal M}_{\underline{\mu} \underline{\nu}}
    =0$.  We then find the new constraint  
${\cal V}^{\underline{\mu}}{\cal N}_{\underline{\mu}} = 0$. 
The null eigenvector can be obtained so that the new constraint is 
written as  
\begin{equation}
\label{chi}
\chi = {\cal N}_{\underline{D-r}} - {\cal M}_{\underline{D-r} \;
\underline{\beta'}}M^{\underline{\beta'} \; \underline{\alpha'}}
{\cal N}_{\underline{\alpha'}}.
\end{equation}
We split the last component of $\underline{\mu}$ as $(\underline{\mu'},
\underline{D-r})$ and from Eq.(\ref{Phi.mu-2}) we get  
\be
\label{lambdanu'}
\lambda^{\underline{\nu'}} = - {\cal M}^{\underline{\mu'} \; \underline{\nu'}}
\left( {\cal N}_{\underline{\mu'}} + {\cal M}_{\underline{\mu'} \;
\underline{D-r}}\lambda^{\underline{D-r}} \right).
\ee
Consistency of $\chi$ then allow us to determine
\be
\label{lambdaD-r}
\lambda^{\underline{D-r}} = - \frac{ \left\{ \chi(\sigma) ,
\frac{1}{2} \varphi(\sigma') \right\} + {\cal
M}^{\underline{\rho'}
\; \underline{\mu'}}{\cal N}_{\underline{\rho'}}\left\{ \chi ,
\Phi_{\underline{\mu'}} \right\} }{ {\cal V}^{\underline{\mu}} \left\{ \chi ,
\Phi_{\underline{\mu}} \right\}}, 
\ee
where 
\be
\label{VmuchiPhi}
{\cal V}^{\underline{\mu}}\{ \chi , \Phi_{\underline{\mu}} \} =
\left\{ \chi , \Phi_{\underline{D-r}} \right\} -
{\cal M}_{\underline{\rho'} \; \underline{D-r}}
{\cal M}^{\underline{\rho'} \; \underline{\mu'}}
\left\{ \chi , \Phi_{\underline{\mu'}} \right\},
\ee
Therefore the
denominator of Eq.(\ref{lambdaD-r}) is non-vanishing. In this way we
have determined all components of the Lagrange multiplier
$\lambda^{\underline{\mu}}$ so that $\Phi_{\underline{\mu}}$ is
second class.  

Since the Hamiltonian has the same form as the relativistic particle
Hamiltonian we find that the Lagrange multiplier $\mu$ vanishes. This
implies then that $\chi$ is a second class constraint. 

We still have to analyze $\{ \varphi_1 , \chi \}$. A short calculation
shows that 
\bea
\label{str:phi1chi}
\{ \varphi_1 , \chi \} &=& \{ {\cal P}_{\mu}x^{'\mu} ,
{\cal V}^{\underline{\mu}} {\cal N}_{\underline{\mu}} \} \nonumber\\
&=& \{ {\cal P}_{\mu} , {\cal V}^{\underline{\nu}} \}x^{'\mu}{\cal
N}_{\underline{\nu}} + \{ x^{'\mu} , {\cal V}^{\underline{\nu}} \} {\cal
P}_{\mu}{\cal N}_{\underline{\nu}} 
+ \{ {\cal P}_{\mu} , {\cal N}_{\underline{\nu}} \} x^{'\mu}{\cal
V}^{\underline{\nu}} + \{ x^{'\mu} , {\cal N}_{\underline{\nu}} \} {\cal
P}_{\mu}{\cal V}^{\underline{\nu}}.
\eea
It is not possible to know whether this vanishes or not since it
depends explicitly on the metric. So if $\{ \varphi_1 , \chi \} \neq
0$ we have an even number of second class constraints. In this
case there are $D + r - 3$ degrees of freedom in phase space. If, on
the other side, $\{ \varphi_1 , \chi \} = 0$ we have an odd
number of second class constraints and again there must be a
combination of them which is first class. Since the 
structure of the constraints and consistency conditions are similar to
those of the relativistic particle the first class constraint will
have the same form as in Eq.(\ref{xi}). Therefore we have the second
class constraints $\Phi_{{\underline{\mu}}^\prime}, \varphi_2$ and $\chi$ and the first
class constraints $\varphi_1, {\cal P}_{(a)(b)}$ and $\xi$. In this case
there are again $D + r - 3$ degrees of freedom in phase space. 
Therefore the degrees of freedom counting is independent of the value of
$\{ \varphi_1 , \chi \}$. 

\section{ISOMETRIES AND FIRST CLASS CONSTRAINTS}
\label{sec5}

Let us consider the role of isometries when the metric is
degenerate. We will start with the non-relativistic particle
case. Consider the following transformation 
\begin{equation}
\label{iso:transf}
\delta x^{i} = - \epsilon^{\alpha}K^{i}_{\alpha},
\end{equation}
where $\epsilon^{\alpha}$ is a constant infinitesimal parameter and
$K_\alpha$, $\alpha=1,\dots,r$ represents a set of $r$ vectors. The
Lagrangian Eq.(\ref{lagrangian}) transforms as
\be
\label{iso:delL2}
\delta L = - \frac{1}{2} m \epsilon^{\alpha} ( K^{l}_{\alpha}
\partial_{l} g_{ij} 
+ g_{lj}\partial_{i}K^{l}_{\alpha} + g_{il}\partial_{j}K^{l}_{\alpha} )
\dot{x}^{i}\dot{x}^{j} - m \dot{\epsilon}^{\alpha}
K^{i}_{\alpha}g_{ij}\dot{x}^{j}.
\ee
Then the transformation Eq.(\ref{iso:transf}) will be an invariance
of the action if the Lie derivative of the metric vanishes
\begin{equation}
\label{iso:isomet}
({\mathcal L}_{K_{\alpha}}g)_{ij} = K^{m}_{\alpha}\partial_{m}g_{ij} +
(\partial_{i}K^{m}_{\alpha})g_{mj} + (\partial_{j}K^{m}_{\alpha})g_{im}=0,
\end{equation}
that is, if the $K_\alpha$ are Killing vectors. Notice that we do not
need the inverse of the metric anywhere so that it applies also to the
case of a degenerate metric. 

Now let us consider the degenerate metric given by $g_{ij} = e_i^a e_j^b
h_{ab}$. According to the analysis made in Section \ref{sec2} it has
$D-r$ null eigenvectors $U^{i}_{\alpha}$ which will imply in the $D-r$
constraints $\Phi_{\alpha} = 0$. These constraints will be first class
if $\partial_{[i}e^{a}_{j]} = 0$ and the transformations generated by
them Eqs.(\ref{local_symmetries}) will leave the Lagrangian
Eq.(\ref{lagrangian}) invariant. However if we consider the
transformations Eqs.(\ref{local_symmetries}) rewritten as 
\be
\label{iso:deltax}
\delta x^{i} = - \epsilon^{\alpha}(t)U^{i}_{\alpha},
\ee
then the invariance of the Lagrangian Eq.(\ref{lagrangian}) requires
that 
\be
\label{iso:varisometria}
U^{m}_{\alpha}\partial_{m}g_{ij} +
(\partial_{i}U^{m}_{\alpha})g_{mj} +
(\partial_{j}U^{m}_{\alpha})g_{im} = 0,
\ee
that is, $({\mathcal L}_{U_{\alpha}}g)_{ij}=0$. Therefore the 
Lagrangian is invariant if there is an isometry of the degenerate
metric with the Killing vectors being the null eigenvectors of the metric. 

Let us remark that the transformation Eq.(\ref{iso:deltax}) is local
in contrast to the non-degenerate case where the parameter does not
depend on time. This is due to the fact that the transformation
Eq.(\ref{iso:deltax}) is generated by a first class constraint which
do not exist for the non-degenerate case. 

Let us now consider the situation for strings. In the locally
integrable case $\partial_{[\mu}e_{\nu]}^{a'} =
0$ we have from Eqs.(\ref{StrMmunu},\ref{104}) that 
\bea
{\cal M}_{\underline\mu \underline\nu}(\sigma) &=& 0, \\
{\cal N}_{\underline\nu}(\sigma) &=& - c ( {\cal
  U}_{\underline\nu}^\mu \partial_{\mu}G_{\lambda \rho} 
x^{'\lambda} x^{'\rho}(\sigma) + 2 \partial_{\sigma}{\cal
  U}_{\underline\nu}^\mu G_{\lambda \mu}x^{'\lambda}(\sigma) ). 
\label{str:Nmured}
\eea
This form of $ {\mathcal N}_{\underline\mu} $ does not allow all
constraints to be first class as discussed in Section
\ref{sec4}. However, if we consider an isometry 
condition in the form 
\be
\label{str:isomet}
{\cal U}^{\mu}_{\underline\nu}\partial_{\mu}G_{\lambda \rho} + G_{\lambda
\mu}\partial_{\rho}{\cal U}^{\mu}_{\underline\nu} + G_{\mu
\rho}\partial_{\lambda} {\cal U}^{\mu}_{\underline\nu} = 0,
\ee
where ${\mathcal U}_{\underline\nu}$ represents a null eigenvector of
$G_{\lambda\mu}$ then ${\cal N}_{\underline\mu}(\sigma) = 0$. Hence 
$\dot{\Phi}_{\underline\mu} = 0 $ does not generate new constraints and we
have a set of first class constraints given by $ \Phi_{\underline\mu} ,\;
\varphi_1 \; e \; \varphi_2 $. Then the theory will have $D - ( D - r
+ 2 ) = r - 2$ degrees of freedom in configuration space. This shows
that the isometry condition is necessary in order to reduce all
constraints to first class ones.

\section{QUANTIZATION OF THE NON-RELATIVISTIC PARTICLE}
\label{sec6}

In this section we will consider the path integral quantization of the
non-relativistic particle, discussed in Section \ref{sec2}, for the
case $D-r$ even. We recall that in this case there are only second
class constraints. To deal with then we will use the Senjanovic
formulation \cite{senjanovic} appropriate to handle second class
constraints in the path integral and the BFT conversion method
\cite{BFT} which turns the second class constraints into a set of
first class constraints satisfying an Abelian algebra.

In the Senjanovic formulation a delta functional of the second class
constraints Eq.(\ref{constraints}) is inserted into the measure of the
path integral as well as the determinant of their Poisson brackets
Eq.(\ref{constraint_algebra}), that is
\be
\label{partition}
Z = \int Dx^{i^\prime} \,\, Dx^\alpha \,\, DP_{i^\prime} \,\,
DP_\alpha \,\, {\det}^\frac{1}{2} M \,\, \delta ( \phi ) \,\, e^{-iS},
\ee
where the action is obtained from the Hamiltonian
Eq.(\ref{hamiltonian})
\be
\label{action}
S = \int dt \,\, \left( P_{i^\prime} \dot{x}^{i^\prime} + P_\alpha
  \dot{x}^\alpha - \frac{1}{2m} \tilde{g}^{i^\prime j^\prime} P_{i^\prime}
  P_{j^\prime} \right). 
\ee
We can now perform the integral in $P_\alpha$ by using the delta
  functional. We can also exponentiate the determinant by introducing
  ghost variables $\eta^\alpha$. The resulting action is
  quadratic in $P_{i^\prime}$ so that it can also be integrated
  out. The final effective action is then 
\be
\label{efective_action}
S_{eff} = - \frac{m}{2} \int dt \,\, \tilde{g}_{i^\prime j^\prime} \left(
  \dot{x}^{i^\prime} + \dot{x}^\alpha e_\alpha^a
  \tilde{e}_a^{i^\prime} - \eta^\alpha M_{\alpha\beta}^{i^\prime}
  \eta^\beta \right) \left(
  \dot{x}^{j^\prime} + \dot{x}^\beta e_\beta^a
  \tilde{e}_a^{j^\prime} - \eta^\gamma M_{\gamma\delta}^{j^\prime}
  \eta^\delta \right).
\ee
When performing the last functional integral we get a further
  determinant in the measure of the path integral
\be
\label{final}
Z= \int Dx^{i^\prime} \,\, Dx^\alpha \,\, D\eta^\alpha
{\det}^\frac{1}{2} \tilde{g} \,\, e^{-iS_{eff}},
\ee
and this determinant is a functional of $x^{i^\prime}$ and $x^\alpha$
so that the coordinates get mixed up and the final form
Eq.(\ref{final}) is not very useful. Maybe there is a geometrical
interpretation for the effective action Eq.(\ref{efective_action}) but
it is not apparent in this form. This of course deserves further
investigation. We still have the right degrees
of freedom: $r$ from $x^{i^\prime}$, $D-r$ from $x^\alpha$ and
$-\frac{1}{2}(D-r)$ from the ghosts $\eta^\alpha$ adding up to
$\frac{1}{2}(D-r)$. 

We now turn to the Abelian BFT conversion method. The details of the
method can 
be found in the original papers \cite{BFT}. To convert the second
class constraints into first class ones we need to introduce new
variables satisfying certain conditions. To each second class
constraint $\phi_\alpha(t)$ we associate 
a new variable $\chi_\alpha(t)$ which has vanishing Poisson bracket
with all other variables and the constraints, and has
non-vanishing Poisson brackets among themselves
\be
\label{bracket}
\{ \chi_\alpha(t), \chi_\beta(t^\prime) \} =
  \omega_{\alpha\beta} \,\, \delta(t-t^\prime),
\ee
where $\omega_{\alpha\beta}$ is antisymmetric in its indices. We also
  introduce new first class constraints denoted by
  $\tilde{\phi}_\alpha$ such that 
\be
\label{expansion}
\tilde{\phi}_\alpha(t) = \phi_\alpha(t) + \sum_{n=1}^\infty
\tilde{\phi}_\alpha^{(n)}(t).
\ee
If we assume that the first term in the sum is of the type 
\be
\label{first_term}
\tilde{\phi}_\alpha^{(1)}(t) = \int dt^\prime \,\, X(t-t^\prime) \,\, 
\chi_\alpha(t^\prime), 
\ee
and that the new constraints form an Abelian algebra we readily
find that $X(t-t^\prime) = \delta(t-t^\prime)$, 
$\omega_{\alpha\beta} = - M_{\alpha\beta}$ and no higher order terms
are needed in Eq.(\ref{expansion}). Then
\bea
\label{new}
\{ \chi_\alpha, \chi_\beta \} &=& - M_{\alpha\beta}, \nonumber \\
\tilde{\phi}_\alpha &=& \phi_\alpha + \chi_\alpha.
\eea

We can now introduce physical variables $\tilde{x}^i, \tilde{P}_i$ which
have vanishing Poisson brackets with the first class constraints
$\tilde{\phi}_\alpha$. Assuming an expansion like Eq.(\ref{expansion})
for each variable we find that only the first correction is
needed. The final result is 
\bea
\label{physical_variables}
\tilde{x}^{i^\prime} &=& x^{i^\prime} + \chi_\alpha M^{\alpha\beta}
e_\beta^a \tilde{e}_a^{i^\prime}, \nonumber \\
\tilde{x}^\alpha &=& x^\alpha + M^{\alpha\beta} \chi_\beta, \nonumber \\
\tilde{P}_{i^\prime} &=& P_{i^\prime} + \chi_\alpha M^{\alpha\beta}
\partial_{i^\prime} ( e_\beta^a \tilde{e}_a^{j^\prime} ) P_{j^\prime},
\nonumber \\
\tilde{P}_\alpha &=& P_\alpha + \chi_\beta M^{\beta\gamma}
\partial_\alpha (e_\gamma^a \tilde{e}_a^{j^\prime} ) P_{j^\prime}. 
\eea
The first class Hamiltonian is then easily found to be
\bea
\label{first_class_hamiltonian}
\tilde{H}_c (x^i, P_i, \chi_\alpha) &=& H_c (\tilde{x}^i, \tilde{P}_i
) \nonumber \\
& = & \frac{1}{2m} \tilde{g}^{i^\prime j^\prime} (\tilde{x}^i)  \left(
  P_{i^\prime} + \chi_\alpha M^{\alpha\beta} 
  \partial_{i^\prime} ( e_\beta^a \tilde{e}_a^{k^\prime} ) P_{k^\prime}
  \right) \left( P_{j^\prime} + \chi_\alpha M^{\alpha\beta}
  \partial_{j^\prime} ( e_\beta^a \tilde{e}_a^{k^\prime} ) P_{k^\prime}
  \right),
\eea
and the Lagrangian in first order form reads
\be
\label{lagrangian_first}
L = P_i \dot{x}^i + \frac{1}{2} \dot{\chi}_\alpha M^{\alpha\beta}
\chi_\beta - \tilde{H}_c - \lambda^\alpha \tilde{\phi}_\alpha.
\ee
The first class constraints give rise to gauge transformations which
are found by computing the Poisson brackets of the variables with the
constraints. The gauge transformations which leave the Lagrangian
Eq.(\ref{lagrangian_first}) invariant are
\bea
\label{gauge_transf}
\delta \tilde{x}^i &=& \delta \tilde{P}_i = 0, \nonumber \\
\delta \chi_\alpha &=& - \Lambda^\beta M_{\alpha\beta},
\eea
where $\Lambda^\alpha(t)$ is the gauge parameter. As expected
$\tilde{x}^i$ and $\tilde{P}_i$ are gauge invariant and only the newly
introduced variables are not gauge invariant. 

Then we were able to transform the second class constraints into a set of
Abelian first class constraints by adding a new variable $\chi_\alpha$
which obeys first order equations of motion and is pure gauge. We can
find physical variables $\tilde{x}^i$ and $\tilde{P}_i$ but the
resulting Hamiltonian and Lagrangian mix the new and old variables. It
was not possible to find a Lagrangian or Hamiltonian solely in terms
of the new variables.  

\section{CONCLUSIONS}
\label{sec7}

We have analyzed the constraint structure for relativistic and
non-relativistic particles and strings in a background described by a
degenerate metric. We have shown that there 
is a rich structure behind such systems and that the resulting
dynamics is non-trivial. A common feature of all cases is the
consistency condition of the form $ N_\alpha + M_{\alpha \beta}
\lambda^\beta = 0$. It allow us to repeat the same procedures in the
classification of the constraints in all cases. 

We have performed a complete classification of the constraints for the
non-relativistic particle. When $e_\mu^a$ is locally integrable all
constraints are first class and the particle has $r$ degrees of
freedom in configuration space. An effective Lagrangian can be derived
in this case. On the other side, when $e_\mu^a$ is
not integrable we can consider two cases depending on whether
$N_\alpha$ vanishes or not. For non-vanishing $N_\alpha$  and $D-r$
even all constraints are second class and there are $\frac{1}{2}(D+r)$
degrees of freedom; if $D-r$ is odd all constraints are still second
class but now there are $\frac{1}{2}(D+r+1)$ degrees of freedom. For
both cases it is possible to find the effective Lagrangian. In
the case of vanishing $N_\alpha$ and $D-r$ even all constraints are
again second class and the number of degrees of freedom is
$\frac{1}{2}(D+r)$; if $D-r$ is odd then all constraints but one are
second class and the remaining one is first class. In this case there
are $\frac{1}{2}(D+r-1)$ degrees of freedom.

For the relativistic particle the constraint analysis follows the same
lines as in the non-relativistic case. We have analyzed only the
$N_{\underline\mu} \not= 0$ case. For $D-r$ even there is one first
class constraint and the remaining ones are second class and the
particle has $\frac{1}{2}(D+r-2)$ degrees of freedom. For $D-r$ odd
all constraints are second class and there are $\frac{1}{2}(D+r-5)$
degrees of freedom. The cases where $N_{\underline\mu}$ vanishes can
be analyzed in the same way as for the non-relativistic particle. 

The same steps can be taken for strings. Again we considered only the
case when ${\cal N}_{\underline\mu} \not= 0$. For $D-r$ even there are
two first class constraints and the remaining ones are second
class. There are $\frac{1}{2}(D+r-4)$ degrees of freedom in
configuration space. When $D-r$ is odd all constraints are second
class and describe $\frac{1}{2}(D+r-3)$ degrees of freedom. 

When $e_\mu^a$ is locally integrable  we can immediately obtain the
first class constraints in the case of the relativistic and
non-relativistic particles. However the same is not true for the case of
strings. At this point the isometries associated to the degenerate
metric showed its relevance since it allowed us to obtain the
corresponding first class constraints. It is also important to notice
that the Killing vectors are  the null eigenvectors of
the degenerate metric and this must have some geometrical meaning
which at moment is beyond our understanding.
The isometries of strings in a gravitational background are 
important for the formulation of dual actions \cite{dual}. In this
description the action obtained from a transformation of the
components of the original metric is classically equivalent to the
initial action. The procedure does
not make use of the inverse metric and so a degenerate metric can be
used to analyze this sort of duality. This is presently under
investigation. 

We also have done an initial study towards the  
quantization of systems in a degenerate metric background. We
have considered the non-relativistic particle with 
$D-r$ even so that only second class constraints are present. We have 
performed the path integral quantization \`a la Senjanovic and
alternatively using
the conversion method of BFT which transforms the second class
constraints into first class ones. The resulting effective
actions are complicated and no geometrical structures can be
seen. This shows that more powerful methods of quantization than those
employed here are necessary. It also shows that it is necessary to get
a deeper geometrical insight for the constraints in order to recast
the effective actions in a more compact form. Work in this direction
is in progress.

\section{ACKNOWLEDGMENTS}

LAC would like to thank FAPESP for financial help. This work was
partially supported by FAPESP and CNPq.


\begin{references}
\bibitem{sing} S.
  W. Hawking and R. Penrose, Proc. Roy. Soc. Lond. {\bf A314}, 529
  (1970); S. W. Hawking and G. F. R. Ellis, The Large Scale Structure
  of Space-Time (Cambridge, 1973); for a recent discussion of the role
  of singularities see G. T. Horowitz and R. Myers,
  Gen. Rel. Grav. {\bf 27} (1995) 915; R. H. Brandenberger,
  ``Nonsingular Cosmology and Planck Scale Physics'',  preprint
  Brown-HET-985, gr-qc/9503001.  
\bibitem{entropy} A. Strominger and E. Martinec, Phys. Lett. {\bf
    B379} (1996) 99; C. G. Callan, Jr. and J. M. Maldacena,
  Nucl. Phys. {\bf B472} (1996) 591; G. T. Horowitz and A. Strominger,
  Phys. Rev. Lett. {\bf 77} (1996) 2368.
\bibitem{ross} G. T. Horowitz and S. F. Ross, JHEP {\bf 9804} (1998)
  015;  A. Lawrence and E. Martinec, 
  Class. Quant. Grav. {\bf 13} (1996) 63. 
\bibitem{horo-marolf} G. T.Horowitz and D. Marolf,
  Class. Quant. Grav. {\bf 12} (1995) 2173.
\bibitem{horowitz} G. T. Horowitz, Class. Quantum Grav. {\bf 8}, 587
  (1991); J. Louko and R. D. Sorkin, Class. Quant. Grav. {\bf
    14} (1997) 179 and references therein.
\bibitem{chamblin} A. Chamblin, ``Topology and Causal Structure'',
  gr-qc/9509046.
\bibitem{7} J. C. Baez, Commun. Math. Phys. 193 (1998) 219.
\bibitem{solowski} L. M. Sokolowski, Acta Phys. Pol. {\bf B6} (1975)
  529, ibidem 657; J. Smallwood, J. Math. Phys. {\bf 20} (1979) 459. 
\bibitem{ashtekar} I. Bengtsson, Class. Quantum Grav. {\bf 8}, 1847
  (1991); T. Jacobson and J. D. Romano, Class. Quantum Grav. {\bf 9},
  L 119 (1992).
\bibitem{9} I. Bengtsson and T. Jacobson, Class. Quant. Grav. 14
  (1997) 3109. 
\bibitem{gauge} A. Ach\'ucarro and P. Townsend, Phys. Lett {\bf B
    180}, 89 (1986); E. Witten, Nucl. Phys. {\bf B 311}, 46 
  (1988); K. Isler and C. Trugenberger, Phys. Rev. Lett. {\bf 63}, 834
  (1989); A. Chamseddine and D. Wyler, Phys. Lett. {\bf B228}, 75
  (1989); D. Cangemi and R. Jackiw, Phys. Rev. Lett. {\bf 69}, 233
  (1992).
\bibitem{witten-top}E. Witten, Commun.. Math. Phys. {\bf 117}, 353 (1988).
\bibitem{strings}O. A. Mattos and V. O. Rivelles, Phys. Rev.
  Lett. {\bf 70}, 1583 (1993).
\bibitem{zaikov} R. P. Zaikov, Phys. Lett. {\bf B263} (1991) 206. 
\bibitem{lindstrom} A. Karlhede and U. Lindstr\"om,
  Class. Quant. Grav. {\bf 3} (1986) L73. 
\bibitem{unge} U. Lindstr\"om and R. von Unge, Phys. Lett. {\bf B403}
  (1997) 233; H. Gustafsson and U. Lindstr\"om, Nucl. Phys. {\bf B540}
  (1999) 520.
\bibitem{marolf} D. M. Marolf, Class. Quant. Grav. {\bf 11} (1994)
  239. 
\bibitem{senjanovic} P. Senjanovic, Ann. Phys. (NY) {\bf 100} (1976)
  227. 
\bibitem{BFT} I. A. Batalin and E. S. Fradkin, Phys. Lett. {\bf B180}
  (1986) 157; Nucl. Phys. {\bf B279}
  (1987) 157; I. A. Batalin and I. V. Tyutin, Intl. J. Mod. Phys. {\bf
  A6} (1991) 3255.
\bibitem{dual} A. Giveon and M. Rocek, ``Introduction to Duality'',
  hep-th/9406178; Nucl. Phys. {\bf B194} (1994) 173; T. Buscher,
  Phys. Lett. {\bf B194} (1987) 59.

\end{references}
\end{document}